\documentclass[12pt]{article}

\title{Topological self-similarity of the random binary-tree model}
\author{Ken Yamamoto and Yoshihiro Yamazaki}
\date{\small
\it Department of physics, Waseda University, Tokyo, Japan\\
\rm e-mail: yamaken@toki.waseda.jp}

\usepackage{amsmath, amssymb, indentfirst, cite}
\usepackage[dvips]{graphicx}

\newcommand{\floor}[1]{\lfloor #1 \rfloor}
\newcommand{\Floor}[1]{\left\lfloor #1 \right\rfloor}

\newcommand{\E}[2]{E_{#1}\left[#2\right]}
\newcommand{\var}[1]{\mathrm{var}\left(#1\right)}

\newcommand{\f}[1]{f\left(#1\right)}

\newcommand{\num}{\nonumber \\}

\begin{document}
\maketitle

\begin{abstract}
Asymptotic analysis on some statistical properties
of the random binary-tree model is developed.
We quantify a hierarchical structure of branching patterns
based on the Horton-Strahler analysis.
We introduce a transformation of a binary tree,
and derive a recursive equation about branch orders.
As an application of the analysis,
topological self-similarity and its generalization
is proved in an asymptotic sense.
Also, some important examples are presented.
\end{abstract}

{\small
topological self-similarity, asymptotic behavior
}

\section{Introduction}
Branching patterns are universal in nature,
including river networks, blood vessels, and dendritic crystals
\cite{Fleury, Ball}.
They usually exhibit intricate forms
(some patterns have been treated
as fractal or multifractal objects).
Branching structures are also fundamental and important tools
for illustrating 
some data structures in computer science \cite{Knuth}
and the classification of species in taxonomy \cite{Schuh}.
For the analysis of branching patterns,
the topological (or graph-theoretic) properties are important
as well as the geometrical ones,
and
even a topological structure is still complicated.

\begin{figure}[tbp]
\centering
\includegraphics[width=.35\textwidth]{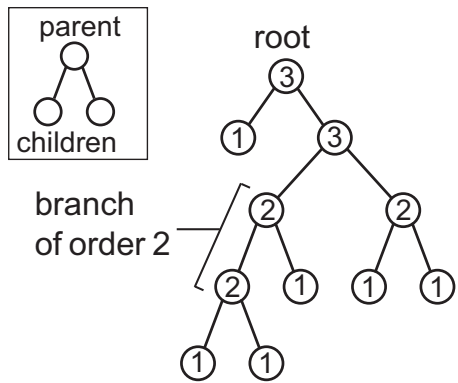}
\caption{A binary tree of magnitude 6.
Nodes are represented by open circles,
and numbers on them are the corresponding
Horton-Strahler indices.}
\label{fig:binarytree}
\end{figure}

The topological structure of a branching pattern is
expressed by a binary tree,
if the pattern is loopless and
all the branching points are two-pronged.
A binary tree can be regarded as a nested structure
of the parent-child relations of nodes (see Fig. \ref{fig:binarytree}).
In order to derive quantitative characteristics
about binary trees,
Horton \cite{Horton} has introduced the idea of branch ordering.
For mathematical convenience,
Horton's method has been refined
by Strahler \cite{Strahler}.
The basic idea of their methods is assignment of numbers,
referred to as Horton-Strahler index,
to the nodes of a binary tree.

Horton-Strahler ordering for a binary tree is defined recursively as follows
(see also Fig. \ref{fig:binarytree}):
(i) each leaf is assigned the order 1,
(ii) a node whose children are both $r$th is assigned $r+1$,
(iii) a node whose children are $r_1$th and $r_2$th ($r_1\ne r_2$) is
	assigned $\max\{r_1,r_2\}$.
In a binary tree,
$r$th {\it branch} is defined as a maximal path connecting $r$th nodes.
The ratio of the number of branches 
between two subsequent orders is called
{\it bifurcation ratio}.
It has been revealed that the bifurcation ratios
become almost constant for different orders
in some actual branching patterns
\cite{Strahler, Feder, Ossadnik, Arenas, Hahn, Binzegger},
which is referred to as {\it topological self-similarity} \cite{Halsey2}.
As a typical instance, many river networks possess
their bifurcation ratios between 3 and 5 irrespective of orders
\cite{Horton, Halsey2}.
The relevance of two types of self-similarity,
`original' self-similarity and topological self-similarity,
has been considered in ramification analysis
\cite{Tokunaga1, Tokunaga2, Vannimenus, Turcotte}.

The number of leaves of a binary tree is called {\it magnitude},
and let $\Omega_n$ denote
the set of topologically different binary trees of magnitude $n$.
The number of the elements of $\Omega_n$ is given by
\begin{equation}
\sharp\Omega_n = \frac{(2n-2)!}{n!(n-1)!} \equiv c_{n-1},
\label{eq:catalan}
\end{equation}
where $c_{n-1}$ is $(n-1)$th Catalan number \cite{Catalan}.
One of the most simple model of a branching structure 
is called random model \cite{Shreve},
where all the binary trees in $\Omega_n$ emerge randomly.
More accurately,
the random model is a probability space $(\Omega_n, P)$,
where $P$ represents the uniform probability measure on $\Omega_n$,
i.e., every binary tree $T\in\Omega_n$ has
the same statistical weight $1/c_{n-1}$.
We denote by $\E{n}{\cdot}$ an average over $\Omega_n$.
We introduce
a random variable $S_{r,n}:\Omega_n\to\mathbb{N}\cup\{0\}$ 
such that
$S_{r,n}(T)$ represents the number of $r$th branches
in a binary tree $T\in\Omega_n$.

The $r$th bifurcation ratio $R_{r,n}$ on $(\Omega_n, P)$ is defined as
\[
R_{r,n} = \frac{\E{n}{S_{r,n}}}{\E{n}{S_{r+1,n}}} \quad (r=1,2,\cdots),
\]
and
topological self-similarity has been confirmed
in the case where magnitude $n$ is sufficiently large.
In fact,
Moon \cite{Moon} has derived
\[
\E{n}{S_{r,n}} = 4^{1-r}n+\frac{1-4^{1-r}}{6}+O(n^{-1}),
\]
and 
\begin{equation}
R_{r,n} =4-\frac{4^r}{2n}+O(n^{-2}) 
	\to 4 \quad (n\to\infty). \label{eq:Horton's law}
\end{equation}
Therefore, the random model is topologically self-similar
in an asymptotic sense,
and the limit value of $R_{r,n}$ is 4.
Moreover, the present authors \cite{Yamamoto} have derived
\begin{align}
\frac{\E{n}{S_{r,n}^k}}{\E{n}{S_{r+1,n}^k}}
&=4^k-\frac{4^{k+r-1}k^2}{2n}+O(n^{-2}) \num
&\to 4^k, \label{eq:momentratio}
\end{align}
and this relation can be regarded
as a generalization of Eq. \eqref{eq:Horton's law}.
Other results on the Horton-Strahler analysis and tree structures
are found in Refs. 
\cite{Yekutieli, Werner, Smart, Meir, Ruskey, Devroye, Prodinger, Toroczkai}.

In the present paper,
we focus on a random variable $f(S_{r,n})$, where
$f:\mathbb{N}\cup\{0\}\to\mathbb{R}\mbox{ (or $\mathbb{C}$)}$
is a certain function
(further assumptions for $f$ are stated later).
We first derive a recursive relation between $\E{n}{f(S_{r,n})}$ and
$\E{m}{f(S_{r-1,m})}$.
Then, we also derive the asymptotic form of $\E{n}{f(S_{r,n})}$, and
show topological self-similarity about $f$
(or simply referred to as
{\it generalized} topological self-similarity),
in the sense that $f$-bifurcation ratio
\[
R_{r,n}^f = \frac{\E{n}{f(S_{r,n})}}{\E{n}{f(S_{r+1,n})}}
\]
is asymptotically independent of $r$.
Cleary, $R_{r,n}^f$ is reduced to $R_{r,n}$ when $f(x)=x$.

\section{Transformation of binary tree} \label{sec:trans}
\begin{figure}[htbp]
\centering
\includegraphics[width = 12cm]{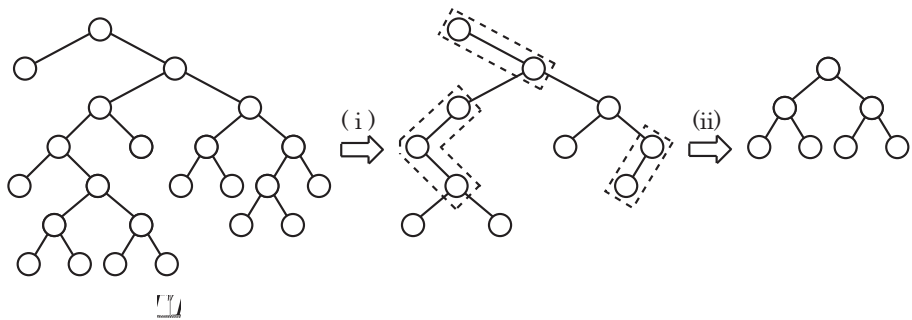}
\caption{An illustration of $\Phi_n$ for $n=12$:
(i) removal of the leaves of $T$, (ii) contraction.}
\label{fig:phi}
\end{figure}

First, we introduce a transformation $\Phi_n$.
For a binary tree $T\in\Omega_n$, a new binary tree $\Phi_n(T)$
is constructed from the following two steps:
(i) remove all the leaves from $T$, % remove or eliminate?
(ii) if a node with only one child appears,
	such a node is merged with its child
	(this operation is called {\it contraction} in the graph theory).
Figure \ref{fig:phi} illustrates these two steps.
The magnitude of $\Phi_n(T)$ is at most $\floor{n/2}$,
because a pair of first-order branches is needed to create
a second-order branch.
Hence,
\[
\Phi_n:\Omega_n\to\bigcup_{m=1}^{\Floor{\frac{n}{2}}}\Omega_m.
\]

We introduce $\Omega_n^m = \Phi_n^{-1}(\Omega_m)$,
which is explicitly expressed as
$\Omega_n^m=\{T\in\Omega_n; S_{2,n}(T)=m\}$.
$\{\Omega_n^m\}_m$ is a partition of $\Omega_n$, that is,
\[
\Omega_n=\bigcup_{m=1}^{\Floor{\frac{n}{2}}}\Omega_n^m,\quad
\Omega_n^m\cap\Omega_n^{m'}=\emptyset\quad (m\ne m').
\]
By definition, we have
$S_{r-1,m}(\Phi_n(T))=S_{r,n}(T)$ if $T\in\Omega_n^m$,
and we regard that this is a relation which connects variables about
two subsequent orders ($r$th and $(r-1)$th).
For example, as for the binary tree $T$
in Fig. \ref{fig:phi} ($n=12$, $m=4$),
we can easily check the following relations:
\[
S_{1,m}(\Phi_n(T)) = 4 = S_{2,n}(T), \quad
S_{2,m}(\Phi_n(T)) = 2 = S_{3,n}(T), \quad
S_{3,m}(\Phi_n(T)) = 1 = S_{4,n}(T).
\]

\begin{figure}[htbp]
\centering
\includegraphics[width=.65\textwidth]{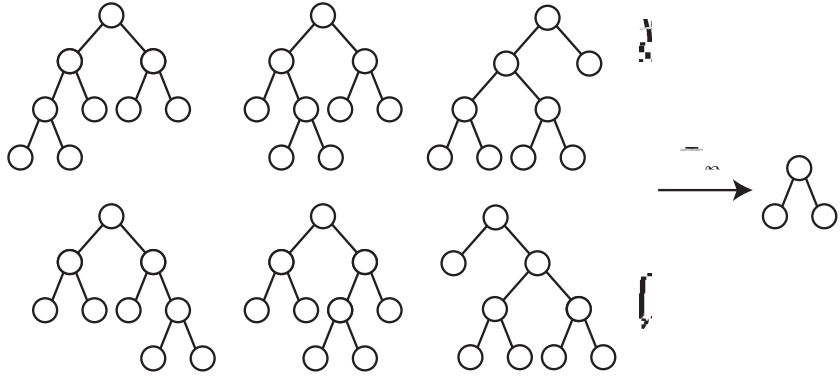}
\caption{Some binary trees are mapped to
the identical binary tree by an operation of $\Phi_n$.
The multiplicity in this case is $\mu_5^2=6$ ($n=5$, $m=2$).}
\label{fig:onetoone}
\end{figure}

Note that the restriction
$\Phi_n|_{\Omega_n^m}:\Omega_n^m\to\Omega_m$ is not one-to-one
(see Fig. \ref{fig:onetoone} for example).
Then,
for a binary tree $\tau\in\Omega_m\ (1\le m\le\floor{n/2})$,
we introduce multiplicity
\[
\mu_n^m(\tau)=\sharp\{T\in\Omega_n\vert\Phi_n(T)=\tau\}
	\equiv\sharp\Phi_n^{-1}(\tau).
\]

\begin{figure}[htbp]
\centering
\includegraphics[width=\textwidth]{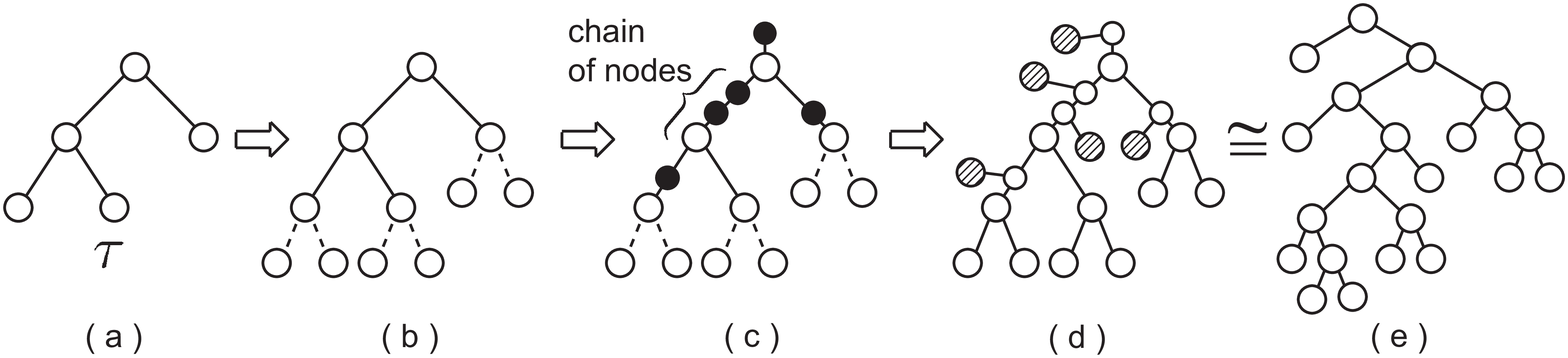}
\caption{An example of $\Phi_n^{-1}(\tau)$ for $n=11$ and $m=3$:
(a) an initial binary tree $\tau\in\Omega_m$,
(b) a pair of nodes is attached to each leaf of $\tau$
	(indicated by the dashed lines),
(c) intermediate nodes (black nodes in the figure) are added,
(d) new leaves (hatched nodes) are attached to the intermediate nodes, and
(e) generated binary tree.
}
\label{fig:inv_phi}
\end{figure}

\begin{figure}[htbp]
\centering
\includegraphics[width=.6\textwidth]{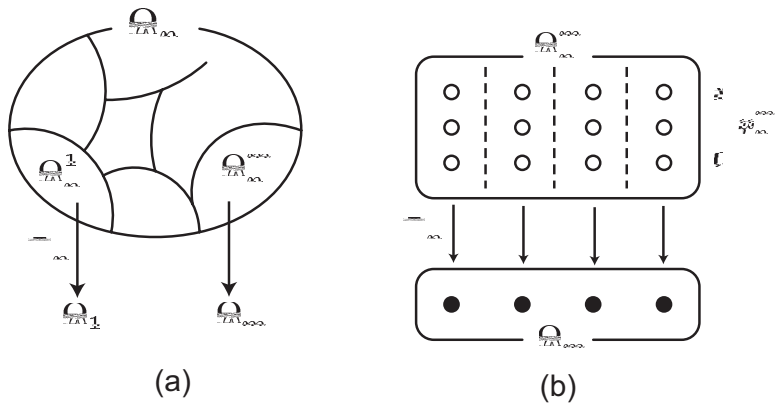}
\caption{Schematical illustration of $\Omega_n^m$, $\Phi_n$,
and $\mu_n^m$.
(a) $\{\Omega_n^m\}_m$ is a partition of $\Omega_n$,
and $\Phi_n(\Omega_n^m)=\Omega_m$.
(b) The open and solid circles represent
	individual binary trees
	(elements of $\Omega_n^m$ and $\Omega_m$, respectively).
	Vertically aligned points in $\Omega_n^m$
	are mapped to
	the identical point in $\Omega_m$.
}
\label{fig:multiple}
\end{figure}

In order to calculate $\mu_n^m(\tau)$,
we trace the inverse process $\Phi_n^{-1}(\tau)$.
As mentioned above, $\Phi_n$ is a removal of the leaves of a binary tree,
and multiplicity $\mu_n^m$ is concerned with a contraction process.
Thus,
the inverse process $\Phi_n^{-1}(\tau)$ can be formed
by attaching $n$ leaves to $\tau$ in the following way.
\begin{enumerate} \renewcommand{\labelenumi}{(\roman{enumi})}
\item A pair of nodes is attached to each leaf of $\tau$.
\item
$n-2m$ `intermediate' nodes are added in the form of a chain,
which is the inverse of contraction.
The number of different ways of adding amounts to $\binom{n-2}{n-2m}$.
\item $n-2m$ leaves are attached to the nodes added in (ii).
Each leaf can be attached independently 
from either left or right, thus
the total number of ways of choosing sides is given by $2^{n-2m}$.
\end{enumerate}
A series of procedures is presented in Fig. \ref{fig:inv_phi}.
From these steps, the total multiplicity $\mu_n^m(\tau)$ is
calculated as
\begin{equation}
\mu_n^m(\tau) = \binom{n-2}{n-2m}2^{n-2m},
\label{eq:mu}
\end{equation}
which depends only on $n$ and $m$, and not on $\tau\in\Omega_m$.
Therefore, 
we hereafter write $\mu_n^m\equiv\mu_n^m(\tau)$ with no confusion.
The connection between $\Omega_n^m$, $\Phi_n$, and $\mu_n^m$ is
depicted in Fig. \ref{fig:multiple}.
From the figure, we can derive the following relation:
\begin{equation}
\mu_n^m=\frac{\sharp\Omega_n^m}{\sharp\Omega_m}.
\label{eq:mu_omega}
\end{equation}
By Eqs. \eqref{eq:catalan}, \eqref{eq:mu}, and \eqref{eq:mu_omega},
the number of the elements of $\Omega_n^m$ is expressed as
\[
\sharp\Omega_n^m=\mu_n^m\cdot\sharp\Omega_m
				= \binom{n-2}{n-2m}2^{n-2m}\frac{(2m-2)!}{m!(m-1)!}
				=\frac{(n-2)!\, 2^{n-2m}}{(n-2m)!\, m!\,(m-1)!}.
\]

Therefore, the average of random variable $f(S_{r,n})$ is expressed as
\begin{align}
\E{n}{f(S_{r,n})} &= \frac{1}{c_{n-1}}\sum_{T\in\Omega_n}f(S_{r,n})(T) \num
&= \frac{1}{c_{n-1}}\sum_{m=1}^{\Floor{\frac{n}{2}}}
	\sum_{T\in\Omega_n^m} f(S_{r,n})(T) \num
&= \frac{1}{c_{n-1}}\sum_{m=1}^{\Floor{\frac{n}{2}}}
	\sum_{T\in\Omega_n^m} f(S_{r-1,m})(\Phi_n(T)) \num
&= \frac{1}{c_{n-1}}\sum_{m=1}^{\Floor{\frac{n}{2}}}
	\mu_n^m \sum_{\tau\in\Omega_m} f(S_{r-1,m})(\tau) \num
&= \frac{1}{c_{n-1}}\sum_{m=1}^{\Floor{\frac{n}{2}}}
	\mu_n^m c_{m-1} \E{m}{f(S_{r-1,m})} \num
&= \frac{n!(n-1)!(n-2)!}{(2n-2)!}\sum_{m=1}^{\Floor{\frac{n}{2}}}
	\frac{2^{n-2m}}{(n-2m)!m!(m-1)!} \E{m}{f(S_{r-1,m})}.
\label{eq:recursive}
\end{align}
Eq. \eqref{eq:recursive} is a recursive relation
about $r$th variable $f(S_{r,n})$ and $(r-1)$th variable $f(S_{r-1,m})$.
The present authors \cite{Yamamoto} have derived
a similar recursive equation
for $\E{n}{S_{r,n}^k}$ and $\E{m}{S_{r-1,m}^k}$.
Compared with the former result,
Eq. \eqref{eq:recursive} is more general and
derivation is much easier.
Yekutieli and Mandelbrot \cite{Yekutieli} have derived that the value
\[
\frac{n!(n-1)!(n-2)!2^{n-2m}}{(2n-2)!(n-2m)!m!(m-1)}
\]
is the probability of finding a binary tree of magnitude $n$
with $m$ branches of order 2.

\section{Asymptotic expansion of $\E{n}{f(S_{r,n})}$}
In this section, we derive the asymptotic form of $\E{n}{f(S_{r,n})}$
by using the recursive equation \eqref{eq:recursive}.
Let us assume that the function $f$ has the following expansion:
\begin{equation}
\E{n}{f(S_{1,n})}\equiv \E{n}{f(n)}\equiv f(n) = a_1n^k+b_1n^{k-1}+O(n^{k-2}).
\label{eq:initial}
\end{equation}
We regard
Eq. \eqref{eq:initial} as the initial condition of
the recursive equation \eqref{eq:recursive}.
We also assume that
$\E{n}{f(S_{r,n})}$ has the form
\begin{equation}
\E{n}{f(S_{r,n})} = a_r n^k+b_r n^{k-1}+O(n^{k-2}),
\label{eq:form}
\end{equation}
where the coefficients $a_r$ and $b_r$ are independent of $n$.

The present authors \cite{Yamamoto}
have already derived the asymptotic form of
the  $k$th moment of $S_{2,n}$ as
\begin{align}
\E{n}{S_{2,n}^k} &=\frac{1}{c_{n-1}}\sum_{m=1}^{\Floor{\frac{n}{2}}}
				\mu_n^m c_{m-1}m^k \num
	&=\left(\frac{n}{4}\right)^k\left(1+\frac{k^2}{2n}\right)+O(n^{k-2}).
\label{eq:moment}
\end{align}
By substituting Eq. \eqref{eq:form} into Eq. \eqref{eq:recursive}
and using Eq. \eqref{eq:moment},
the average of $f(S_{r,n})$ can be calculated as
\begin{align}
\E{n}{f(S_{r,n})}
&= \frac{1}{c_{n-1}}\sum_{m=1}^{\Floor{\frac{n}{2}}}
	\mu_n^m c_{m-1} \left(a_{r-1}m^k+b_{r-1}m^{k-1}+O(m^{k-2})\right) \num
&= a_{r-1}\E{n}{S_{2,n}^k}+b_{r-1}\E{n}{S_{2,n}^{k-1}}+O(n^{k-2}) \num
&= a_{r-1}\left(\frac{n}{4}\right)^k\left(1+\frac{k^2}{2n}\right)
	+b_{r-1}\left(\frac{n}{4}\right)^{k-1}+O(n^{k-2}) \num
&= \frac{a_{r-1}}{4^k} n^k
	+ \left(\frac{b_{r-1}}{4^{k-1}}+\frac{k^2a_{r-1}}{2\cdot 4^k}\right)n^{k-1}
	+O(n^{k-2})
\label{eq:E}
\end{align}

Comparing $O(n^k)$ terms of Eqs. \eqref{eq:form} and \eqref{eq:E},
we get a recursive equation about $\{a_r\}$:
\[
a_r = \frac{a_{r-1}}{4^k},
\]
and the solution $a_r$ is
\begin{equation}
a_r = \left(\frac{1}{4^k}\right)^{r-1}a_1.
\label{eq:a}
\end{equation}
Similarly, $O(n^{k-1})$ terms yield an equation about $\{b_r\}$:
\[
b_r = \frac{b_{r-1}}{4^{k-1}}+\frac{k^2a_{r-1}}{2\cdot 4^k}
	= \frac{b_{r-1}}{4^{k-1}}+\frac{k^2a_1}{2}\frac{1}{4^{kr}}.
\]
The solution of this equation is given by
\begin{equation}
b_r=\left(\frac{1}{4^{k-1}}\right)^{r-1}b_1
	+\frac{k^2a_1}{4^k}\frac{4^{r-1}-1}{6}.
\label{eq:b}
\end{equation}
We note that the general solution of the recursive equation
$
x_{r+1}=sx_r+t\cdot u^r
$
 $(s\ne u)$ 
is given by
\[
x_r=s^{r-1}x_1+tu\frac{u^{r-1}-s^{r-1}}{u-s},
\]
and
in this case we set $s=\frac{1}{4^{k-1}}$, $t=\frac{k^2a_1}{2}$,
and $u=\frac{1}{4^k}$.

Substituting Eqs. \eqref{eq:a} and \eqref{eq:b} into Eq. \eqref{eq:E},
one can obtain
\begin{equation}
\E{n}{f(S_{r,n})}=\left(\frac{n}{4^{r-1}}\right)^k
				\left\{a_1+\frac{1}{n}
					\left(4^{r-1}b_1+\frac{4^{r-1}-1}{6}k^2a_1\right)\right\}
				+O(n^{k-2}).
\label{eq:expansion}
\end{equation}
Eq. \eqref{eq:expansion} is the asymptotic expansion of $\E{n}{f(S_{r,n})}$.

A similar formula can be derived for a $p$-variable function
$f(S_{1,n},S_{2,n},\cdots,S_{p,n})$.
For simplicity,
we introduce the notation
$\f{S_{r,n}^{(p)}}\equiv f(S_{r,n},S_{r+1,n},\cdots,S_{r+p-1,n})$.
Assuming the following asymptotic form
\begin{equation}
\E{n}{\f{S_{1,n}^{(p)}}} = a_1n^k+b_1n^{k-1}+O(n^{k-2}),
\label{eq:initial_multi}
\end{equation}
then
the asymptotic form of $\E{n}{f(S_{r,n}^{(p)})}$
is expressed as
\begin{equation}
\E{n}{\f{S_{r,n}^{(p)}}}
		=\left(\frac{n}{4^{r-1}}\right)^k
			\left\{a_1+\frac{1}{n}
				\left(4^{r-1}b_1+\frac{4^{r-1}-1}{6}k^2a_1\right)\right\}
			+O(n^{k-2}).
\label{eq:expansion_multi}
\end{equation}

\section{Generalized topological self-similarity}
Using the asymptotic formulas
\eqref{eq:expansion} and \eqref{eq:expansion_multi},
we can easily show generalized topological self-similarity.
The asymptotic form of a generalized bifurcation ratio $R_{r,n}^f$
is calculated as
\begin{align}
R_{r,n}^f &\equiv\frac{\E{n}{\f{S_{r,n}^{(p)}}}}{\E{n}{\f{S_{r+1,n}^{(p)}}}}
=4^k-\frac{4^{k+r-1}(6b_1+a_1k^2)}{2a_1n}+O(n^{-2}) \num
&\to 4^k \quad \mbox{as } n\to\infty. \label{eq:similarity}
\end{align}
Therefore,
on the random binary-tree model,
topological self-similarity about $f$
is concluded in an asymptotic sense,
if $f$ has an expansion
as in Eqs. \eqref{eq:initial} or \eqref{eq:initial_multi}.
Note that the limit value of $R_{r,n}^f$ depends only on
the dominant order $k$ of $f$.

Here, we provide several examples.
\begin{enumerate}
\item % example 1
We start from
$f(n)=n^k$ ($a_1=1$, $b_1=0$), and
$\E{n}{f(S_{r,n})}=\E{n}{S_{r,n}^k}$ is the $k$th moment of $S_{r,n}$.
In this case, Eq. \eqref{eq:expansion} is reduced to
\[
\E{n}{S_{r,n}^k}=\left(\frac{n}{4^{r-1}}\right)^k
\left(1+\frac{4^{r-1}-1}{6n}k^2\right)+O(n^{k-2}),
\]
and the asymptotic form of $R_{r,n}^f$ is 
\[
R_{r,n}^f = 4^k-\frac{4^{k+r-1}k^2}{2n}+O(n^{-2}).
\]
Thus, Eq. \eqref{eq:momentratio} is rederived.

\item %example 2
We next consider the asymptotic property
of the variance of $S_{r,n}$.
By the definition $S_{1,n}\equiv n$,
$\var{S_{1,n}}=0$ is easily obtained.
The analytical expression of the variance of $S_{2,n}$ is given by
\begin{equation}
\var{S_{2,n}}\equiv\E{n}{(S_{2,n}-\E{n}{S_{2,n}})^2}
=\frac{n(n-1)(n-2)(n-3)}{2(2n-3)^2(2n-5)}
=\frac{n}{16}-\frac{1}{32}+O(n^{-1}),
\label{eq:initial ex2}
\end{equation}
which has been obtained by Werner \cite{Werner}.
We think that $\var{S_{1,n}}=0$ is exceptional.
Regarding Eq. \eqref{eq:initial ex2} as the initial condition
of calculation
($a_1=\frac{1}{16}$, $b_1=-\frac{1}{32}$, and $k=1$),
the asymptotic form of $\var{S_{r,n}}$ is calculated as
\[
\var{S_{r,n}}=\frac{n}{4^r}-\frac{1}{48}-\frac{1}{6\cdot4^r}+O(n^{-1}).
\]
Therefore,
for sufficiently large $n$, the variance $\var{S_{r,n}}$
decreases almost exponentially with an increase of $r$.

\item %example 3
We next deal with a two-variable function
$f(S_{1,n},S_{2,n})=S_{2,n}/S_{1,n}(=S_{2,n}/n)$.
According to the result
\[
\E{n}{S_{2,n}}=\frac{n(n-1)}{2(2n-3)}
\]
obtained by Werner \cite{Werner},
the initial condition \eqref{eq:initial_multi}
in this case is calculated as
\[
\E{n}{\frac{S_{2,n}}{S_{1,n}}}=
\E{n}{\frac{S_{2,n}}{n}}=\frac{\E{n}{S_{2,n}}}{n}
=\frac{n-1}{2(2n-3)}
=\frac{1}{4}+\frac{1}{8n}+O(n^{-2}).
\]
Thus, we have $a_1=\frac{1}{4}$, $b_1=\frac{1}{8}$ and $k=0$,
and Eq. \eqref{eq:expansion_multi} yields
\[
\E{n}{\frac{S_{r+1,n}}{S_{r,n}}}=\frac{1}{4}+\frac{4^{r-2}}{2n}+O(n^{-2}).
\]
On the other hand, by using Eq. \eqref{eq:Horton's law},
\[
\frac{\E{n}{S_{r+1,n}}}{\E{n}{S_{r,n}}}=\frac{1}{R_{r,n}}
	=\frac{1}{4}+\frac{4^{r-2}}{2n}+O(n^{-2}).
\]
In conclusion, we obtain
\[
\E{n}{\frac{S_{r+1,n}}{S_{r,n}}}=\frac{\E{n}{S_{r+1,n}}}{\E{n}{S_{r,n}}},
\]
if $O(n^{-2})$ terms are neglected.
This relation is quite simple in appearance,
but it is nontrivial.

\end{enumerate}

\section{Discussion}
The random binary-tree model is highly simplified model
and it seems not so physical
in the sense that it is not directly related to actual patterns.
However,
actual branching patterns are usually affected stochastic effects, and
the randomness is incorporated also in the random model.
Advantage of the random binary-tree model is that
analytical calculations can be widely performed,
and that some of such calculations explain
properties of actual patterns.
Thus, the random model is important
as a prototype of branching systems.
In addition, another significance of the random model is
concerned with statistical mechanics.
A branching system with fluctuations
can be regarded as a statistical ensemble,
and
each binary tree in $\Omega_n$ represents a microscopic state.
From this point of view,
the random model $(\Omega_n, P)$ is regarded
as the microcanonical ensemble.
(In fact,
the uniform measure $P$ corresponds to the principle of equal weight.)
Therefore, the random model is important
for the theoretical foundation of
the statistical physics of branching systems.

In the present paper,
the asymptotic form of $\E{n}{f(S_{r,n})}$ and
generalized topological self-similarity
are confirmed asymptotically for a wide class of $f$.
We only assume that $f$ has an expression as in Eq. \eqref{eq:initial},
which is
a Laurent expansion of $f$ around infinity.
Hence, Eq. \eqref{eq:initial} is valid
if $f$ does not have an essential singularity at infinity.
Polynomial and rational functions are typical examples
of such functions,
and as shown in the previous section,
important random variables on the Horton-Strahler analysis are
mostly polynomial or rational functions of $S_{r,n}$.

We proved topological self-similarity about $f$ on the random model,
and
we expect that such a generalized topological self-similarity
is also valid for some actual branching patterns.
For example, topologically self-similar patterns,
such as river networks, are expected
to be topologically self-similar about $f$, for some class of $f$.
We need further observational, experimental, and numerical
researches for the solution of this problem.

The random binary-tree model is a graph-theoretic model,
where geometrical properties are all neglected.
As a refinement of the Horton-Strahler analysis,
ramification analysis \cite{Tokunaga1, Tokunaga2, Tokunaga3, Tarboton}
describes how many side-branches emerge.
Ramification analysis is still a topological model,
but it is related to a fractal structure.
Based on the methods and results in this paper,
we expect that
some asymptotic properties of random variables
in ramification analysis are obtained,
and that the
more profound comprehension of
a connection between
topological self-similarity and original self-similarity
can be obtained.

\section{Conclusion}
We have first introduced the transformation $\varPhi_n$
in Sec. \ref{sec:trans},
and recursive equation \eqref{eq:recursive} is obtained.
Eq. \eqref{eq:recursive} can be solved asymptotically,
if $f(n)$ is expressed as in Eq. \eqref{eq:initial}.
Solution \eqref{eq:expansion} is the asymptotic form
of $\E{n}{f(S_{r,n})}$.
A similar result \eqref{eq:expansion_multi} is derived for
a multivariable function.
Topological self-similarity about $f$
is confirmed in Eq. \eqref{eq:similarity}.
We have also presented some calculations as examples.

\end{document}